\documentclass{anabs}
\usepackage{graphicx}
\usepackage{times}
\Volume{1-2}              
\Year{2003}              
\Month{02}               
\Pagespan{000}{000}      

\begin{document}
\lhead[\thepage]{A.N. Author: Title}
\rhead[Astron. Nachr./AN~{\bf 324} (2003) 1/2]{\thepage}
\headnote{Astron. Nachr./AN {\bf 324} (2003) 1/2, 000--000}

\title{Initial Optical Results for the ChaMPlane Survey}

\author{Ping Zhao\inst{1}, Jonathan Grindlay\inst{1}, Peter Edmonds\inst{1},
Jaesub Hong\inst{1}, Johnathan Jenkins\inst{1}, Eric Schlegel\inst{1},
Haldan Cohn\inst{2} \and Phyllis Lugger\inst{2}}
\institute{Harvard-Smithsonian Center for Astrophysics, 60 Garden St.,
Cambridge, MA 02138 U.S.A.  \and Department of Astronomy, Indiana
University, Swain Hall West, Bloomington, IN 47405 U.S.A.}

\correspondence{zhao@cfa.harvard.edu}

\maketitle

\section{Introduction}
We present initial, and representative, optical results from the
Chandra Multiwavelength Plane (ChaMPlane) Survey (see paper by
Grindlay {\it et al.} in this Volume).  ChaMPlane is a project to
identify a large sample of serendipitous X-ray sources in deep
galactic plane fields imaged by the Chandra X-ray Observatory in order
to determine the populations of accretion-powered binaries in the
Galaxy.  The primary goal of ChaMPlane is to identify Cataclysmic
Variables (CVs) and quiescent Low Mass X-ray Binaries (qLMXBs), in
order to constrain and ultimately measure their number and space
density luminosity functions.  The secondary objectives are to
determine the Be High-Mass X-ray Binary (BeHMXB) content and stellar
coronal source density in the Galactic Plane.

\section{Observations and Results}
The ChaMPlane Optical Survey is one of NOAO's Long Term Survey
Programs.  In the past two years we have successfully conducted the
ChaMPlane survey and obtained deep images in the V, R, I and H$\alpha$
bands for 24 Chandra fields from CTIO (17 fields) and KPNO (7 fields)
4-m telescopes with the Mosaic ($36^{\prime}\times36^{\prime}$)
camera, with a total coverage of more than eight square degrees in the
galactic plane.  This is the deepest H$\alpha$ survey on portions of
the galactic plane; and it covers $\sim$5$\times$ the area of the
enclosed ChaMPlane X-ray Survey.  The ChaMPlane NOAO imaging survey
serves two purposes: 1) to provide images for identifying the optical
counterparts of Chandra sources and to measure their V, R, I,
H$\alpha$ magnitudes to enable approximate spectral classification and
constraints on reddening; 2) to identify H$\alpha$ emission objects
(by comparing their magnitudes in H$\alpha$ vs. R filters) as well as
to measure their V, R, and I magnitudes.  The NOAO Survey will
eventually yield $\sim$100 fields (Mosaic) near the galactic plane, or
about 36 square degrees.

\begin{figure}
\resizebox{\hsize}{!}
{\includegraphics*[trim=25 0 0 18]{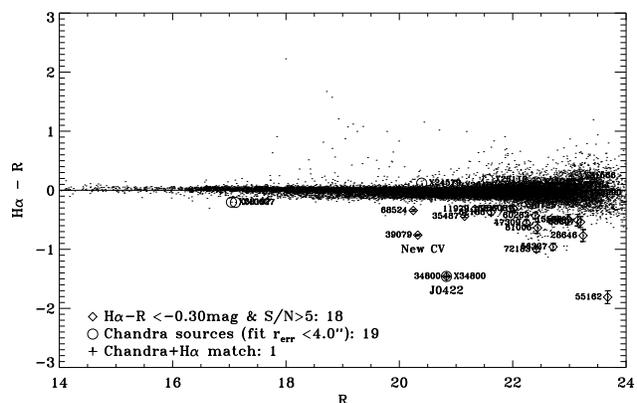}}
\vspace{-0.3in}

\caption{GRO J0422+32 field (H$\alpha -$R) vs.~R color-magnitude
diagram.  Source \#39079 is the first CV candidate found (see Fig.2).}
\label{label1}
\end{figure}

\begin{figure}
  \centering
  \begin{minipage}[c]{0.25\textwidth}
    \centering\includegraphics*[trim=40 0 0 20,scale=0.325]{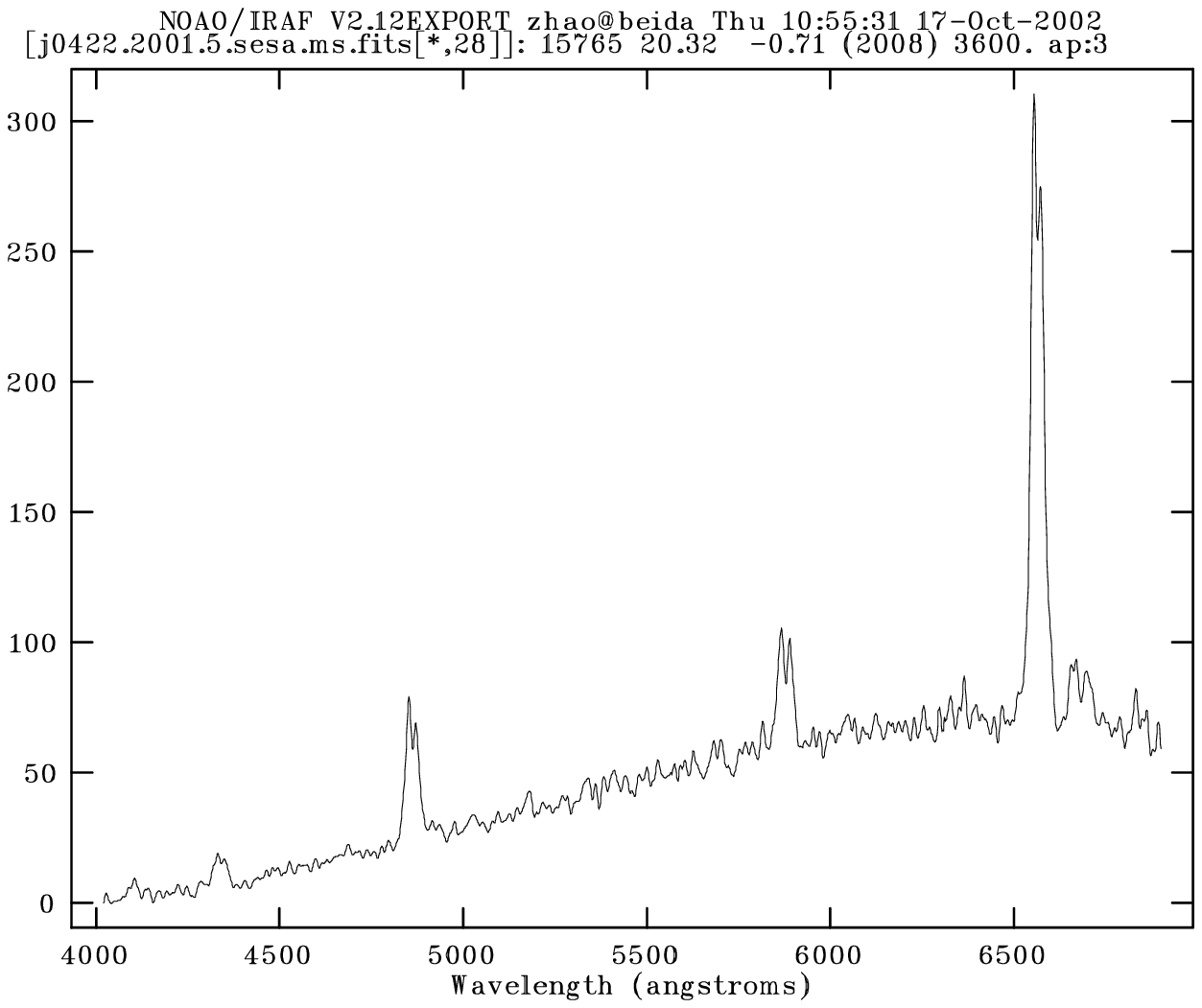}
  \end{minipage}%
  \centering
  \begin{minipage}[c]{0.25\textwidth}
    \centering\includegraphics*[trim=40 0 0 20,scale=0.325]{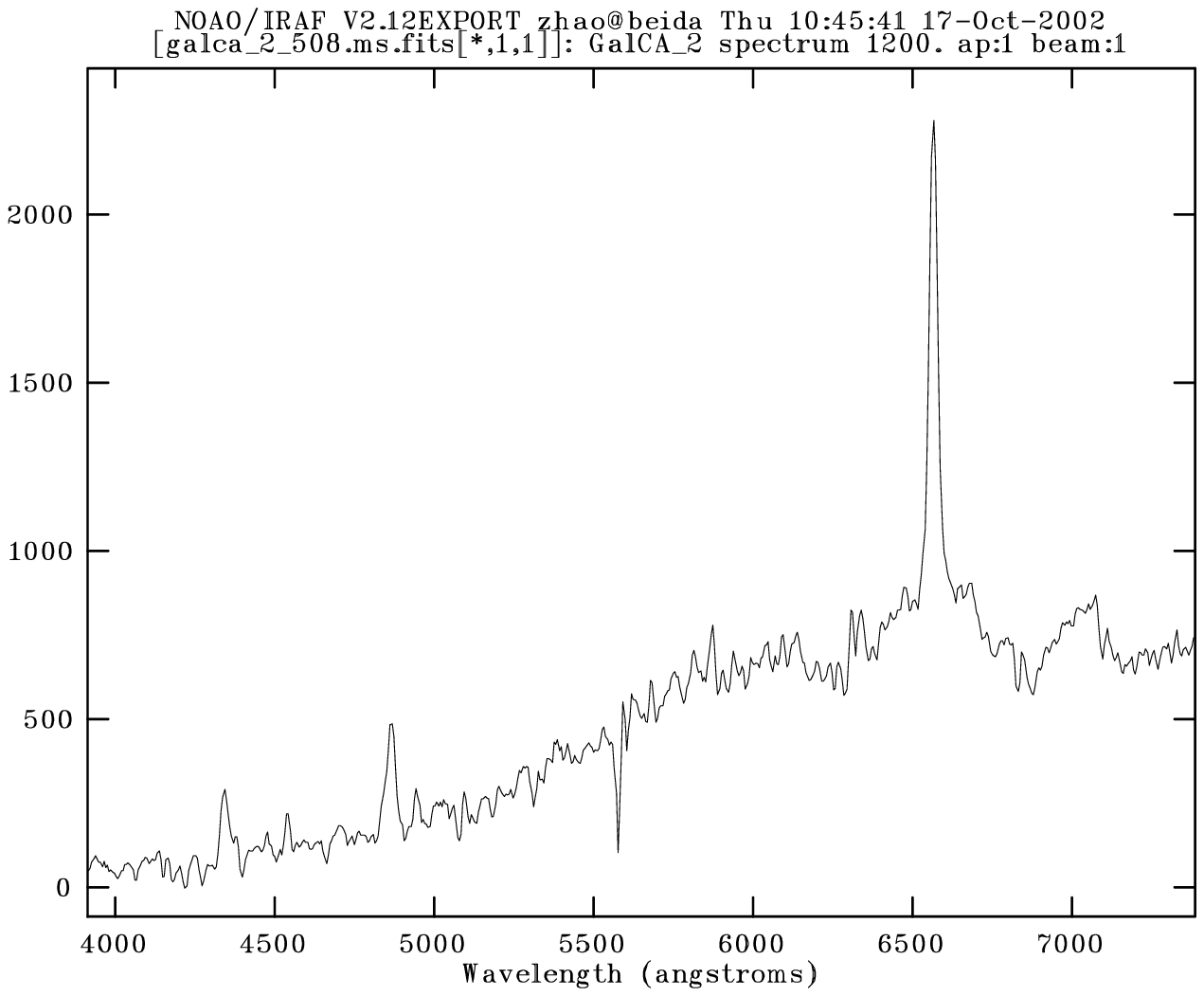}
  \end{minipage}%
\vspace{-0.1in}

\caption{Spectra of two new CV candidates discovered in the ChaMPlane
Survey.  On the left is a WIYN spectrum of source \#39079 in the J0422
field, but outside the Chandra FoV.  It clearly shows double-peaked
H$\alpha$, H$\beta$, H$\gamma$, and He I 5875\AA, 6678\AA\ lines.  On
the right is a Magellan spectrum of the first X-ray-CV candidate, from
the Galactic Center Arc field.  It shows the typical H-Balmer series
and He I lines.}
\label{label2}
\vspace{-0.18in}
\end{figure}

So far the ChaMPlane survey has detected several thousand
serendipitous X-ray sources.  A significant fraction (more than 50\%
for moderately reddened fields) of their optical counterparts are
found in the Mosaic images.  CVs and qLMXBs are identified by their
ubiquitous H$\alpha$ excess as ``blue'' objects in the (H$\alpha -$ R)
vs.~R diagram (see Fig.1).  Spectroscopic follow-ups have been
conducted at the WIYN (3.5-m) and Magellan (6.5-m) telescopes on 12 of
the 24 fields for initial classification and identification of the
Chandra optical counterparts and H$\alpha$ emission sources, including
some H$\alpha$ objects outside the Chandra FoV.  Several new CVs,
including the first X-ray-CV (see Fig.2), Be-HMXB, and many new AGNs
have been found.

This work is supported in part by Chandra grants AR1-2001X, 
AR2-3002A and NSF grant AST-0098683.

\end{document}